\documentclass[12pt,thmsa]{article}
\usepackage{amssymb}

\usepackage{graphicx}

\begin{document}

\author{Alfredo B. Henriques \\
Dep. de F\'{i}sica/Centra\\
Instituto Superior T\'{e}cnico \\
Av. Rovisco Pais, 1049-001 Lisboa, Portugal\\
email: alfredo@fisica.ist.utl.pt}
\title{Loop quantum cosmology\\
and the Wheeler-De Witt equation}
\maketitle

\begin{abstract}
We present some results concerning the large volume limit of loop quantum
cosmology in the flat homogeneous and isotropic case. We derive the
Wheeler-De Witt equation in this limit. Looking for the action from which
this equation can also be obtained, we then address the problem of the
modifications to be brought to the Friedman's equation and to the equation
of motion of the scalar field, in the classical limit.
\end{abstract}

I. \underline{Introduction}.

In recent times we have witnessed an increasing interest in the aplications
of the ideas of loop quantum gravity (lqg) to the problems of cosmology.
This started with a series of seminal papers by M. Bojwald [1-8] on loop
quantum cosmology, where he obtained a number of impressive results, among
them the possibility of removing, in a natural way, the presence of the
cosmological singularity.

Although importing from loop quantum gravity the techniques and ideas, a
number of issues still remain unclear. Loop quantum cosmology (lqc) begins
with a drastic reduction of the phase space of the full theory, resulting,
among other things, in contrast to the situation in the full theory, in a
quantization ambiguity in the form of a parameter which, following [9], we
shall call $\mu _{0}$, playing the role of regulator. The value of $\mu _{0}$
canot be calculated within lqc, but has to be taken from the full theory,
where it signals the fundamental discreteness of geometry, one of the most
fascinating results obtained by loop quantum gravity. Other problems
involve, for instances, the interpretation of the notion of quantum state
and the issue of the time variable, which variable in lqc is replaced by the
momentum variable controlling the evolution of the discrete quantum
constraint equation (see references 10 and 11 for further comments on lqc).

Another important ambiguity arises, as a result of the reduction in the
phase space of the theory. It is well known that the classical hamiltonian of general relativity can be written as the sum of two terms which, in the homogeneous and isotropic case, become proportional to each other; the hamiltonian reduces itself, then, to one term. Using this property, and then quantizing, a discrete equation for the evolution of the system is obtained. This is the procedure normally used in the applications [9]. However, if we separately apply, to the two terms, the same quantization procedure, with the same connection and holonomy, we obtain instead equation (20) below (see also ref. [7]). Thus, it is not indifferent the order followed to derive the evolution equation. In addition, the two equations lead to different large volume limits (section III).

An important part of the work done on lqc has turned around the study of the
semi-classical behaviour of the solutions of the constraint equation, for
different matter hamiltonians, in the large volume limit. This will also be
the focus of our present work. Within the context of a homogeneous and
isotropic universe, assumed to be dominated by a massive scalar field, we
derive a modified form of the Wheeler-De Witt equation from the large volume
limit of lqc, starting from equation (20). We then address the issue of the
modifications to bring to the Friedman's equation and to the equation of
motion of the scalar field, equations usually applied in lqc, by looking for
the action from which, using conventional methods, we derive the above
modified form of the Wheeler-De Witt equation.

\bigskip

II. \underline{The quantum evolution equation}.

IIa. The hamiltonian for the scalar field.

Given the importance of scalar fields in modern cosmology, we derive in this
section the hamiltonian for the scalar field, in a flat homogeneous and
isotropic universe, with the help of the expressions for the connection and
holonomy given in [9].

The holonomy is

\begin{equation}
h_{i}=\cos \frac{\mu c}{2}+2\sin \frac{\mu c}{2}\tau _{i},  \label{1}
\end{equation}
where $\mu $ defines a length along which we integrate, not a physical
length, and c is the real variable containing the information on the
connection. The basis vectors, defining the eigenstates of the momentum
operator $\hat{p},$ are, in the bra-ket notation,

\begin{equation}
<c|\mu >=\exp \frac{i\mu c}{2}  \label{2}
\end{equation}
with the operator

\begin{equation}
\hat{p}=-i\frac{\gamma l_{pl}^{2}}{3}\frac{d}{dc}  \label{3}
\end{equation}
acting as

\begin{equation}
\hat{p}|\mu >=\frac{\gamma l_{pl}^{2}}{6}\mu |\mu >\equiv p_{\mu }|\mu >,
\label{4}
\end{equation}
$\mu \in (-\infty ,+\infty )$ and $l_{pl}^{2}=8\pi G$. The eigenvalues $%
V_{\mu }$ of the volume operator are obtained from the relation $V=|p|^{3/2}$
: 
\begin{equation}
V_{\mu }=(\frac{\gamma l_{pl}^{2}}{6}|\mu |)^{3/2}.  \label{5}
\end{equation}
In the expressions for $p_{\mu }$ and $V_{\mu },$ the constant $\gamma $
represents the Barbero-Immirzi parameter.

We apply these expressions to the hamiltonian operator derived by Thiemann
in [12], which, in the case of a real scalar field, becomes, after the
appropriate calculation of the traces:

\begin{equation}
\hat{H}_{\phi }\psi _{\mu }(\phi )=\{\frac{8m_{pl}}{81l_{pl}^{9}}\frac{9}{4}%
(V_{\mu +\mu _{0}}^{1/2}-V_{\mu -\mu _{0}}^{1/2})^{6}\hat{p}_{\phi }^{2}+%
\frac{m_{pl}}{l_{pl}^{3}}V_{\mu }W(\phi )\}\psi _{\mu }(\phi ).  \label{6}
\end{equation}
The scalar field $\phi $ and the momentum operator $\hat{p}_{\phi }=-i 
\rlap{\protect\rule[1.1ex]{.325em}{.1ex}}h%
d/d\phi $ have been rescaled and are dimensionless; we can check that $\hat{H%
}_{\phi }$ has the correct dimensions of a mass. In the case of a scalar
field of mass $m$, we have the potential $W(\phi )=\frac{1}{2}m^{2}\phi ^{2}$
($m$ in planck units).

We see that the operator $\hat{H}_{\phi }$ naturally selects the ambiguity
parameters, entering into the definition of the inverse scale factor, as $%
l=3/4$ and $j=1/2$.

Writing, as in ref. [9], the physical state as the sum 
\begin{equation}
(\Psi |=\sum_{\mu }\psi _{\mu }(\phi )<\mu |,  \label{7}
\end{equation}
where $\phi $ represents the matter (scalar field of mass m) degrees of
freedom, we derive the action of the hamiltonian on $\psi _{\mu }(\phi )$.
With a potential proportional to $\phi ^{2}$, we may look for solutions in
the form of linear combinations of the Hermite special functions. Taking the
simplest case, 
\begin{equation}
\psi _{\mu }(\phi )=\psi _{\mu }\exp (-\alpha _{\mu }\phi ^{2})  \label{8}
\end{equation}
we find that 
\begin{equation}
\hat{H}_{\phi }\psi _{\mu }(\phi )=E_{\mu }\psi _{\mu }(\phi )  \label{9}
\end{equation}
where 
\begin{equation}
E_{\mu }=\frac{1}{2}\sqrt{A_{\mu }B_{\mu }}m
\rlap{\protect\rule[1.1ex]{.325em}{.1ex}}h%
\label{10}
\end{equation}
and 
\begin{equation}
\alpha _{\mu }=\frac{1}{2}\sqrt{B_{\mu }/A_{\mu }}m 
\rlap{\protect\rule[1.1ex]{.325em}{.1ex}}h%
\label{11}
\end{equation}
with the expressions for $A_{\mu }$ and $B_{\mu }$ taken from (6): 
\begin{equation}
A_{\mu }=\frac{4m_{pl}}{9l_{pl}^{9}}(V_{\mu +\mu _{0}}^{1/2}-V_{\mu -\mu
_{0}}^{1/2})^{6}  \label{12}
\end{equation}
and 
\begin{equation}
B_{\mu }=\frac{m_{pl}}{l_{pl}^{3}}V_{\mu }.  \label{13}
\end{equation}

With these definitions we rewrite $\hat{H}_{\phi }$ in the form 
\begin{equation}
\hat{H}_{\phi }=\frac{1}{2}A_{\mu }\hat{p}_{\phi }^{2}+B_{\mu }\frac{1}{2}%
m^{2}\phi ^{2}.  \label{14}
\end{equation}
In section III. we shall apply these expressions to integrate the Wheeler-De Witt equation, using numerical methods.

\pagebreak

IIb. The discrete evolution equation obtained from the gravitational
operator.

To obtain this equation we shall apply the holonomy operator (1) to the full
hamiltonian operator derived by Thiemann (see Thiemann's [13] and also the
report by A. Ashtekar and J. Lewandowski [14]), defining the regulated
constraint operator: 
\begin{equation}
(\Psi |\hat{H}_{R\varepsilon }(N)=(\Psi |\sum_{\square }(\sqrt{\gamma }\hat{C%
}_{\square }^{eucl}(N)-2(1+\gamma ^{2})\hat{T}_{\square }(N)),  \label{15}
\end{equation}
where the euclidean scalar constraint is 
\begin{equation}
\hat{C}_{\square }^{eucl}(N)=-\frac{iN}{l_{pl}^{4}\gamma ^{3/2} 
\rlap{\protect\rule[1.1ex]{.325em}{.1ex}}h%
}\sum_{i,J}\epsilon ^{ijk}Tr((\hat{h}_{\alpha _{ij}}^{\mu _{0}}-\hat{h}%
_{\alpha _{ji}}^{\mu _{0}})(\hat{h}_{k}^{\mu _{0}})^{-1}[h_{k}^{\mu _{0}},%
\hat{V}])  \label{16}
\end{equation}
and the operator corresponding to the extrinsic curvature terms

\begin{equation}
\hat{T}_{\square }(N)=\frac{iN}{l_{pl}^{8}\gamma ^{3} 
\rlap{\protect\rule[1.1ex]{.325em}{.1ex}}h%
^{3}}\epsilon ^{ijk}Tr((\hat{h}_{i}^{\mu _{0}})^{-1}[\hat{h}_{i}^{\mu _{0}},%
\hat{K}^{\prime }](\hat{h}_{j}^{\mu _{0}})^{-1}[\hat{h}_{j}^{\mu _{0}},\hat{K%
}^{\prime }](\hat{h}_{k}^{\mu _{0}})^{-1}[\hat{h}_{k}^{\mu _{0}},\hat{V}]);
\label{17}
\end{equation}
the smearing function $N$ will be put equal to 1 throughout and the
extrinsic curvature operator $\hat{K}^{\prime }$ expressed as a commutator
[13, 14] 
\begin{equation}
\hat{K}^{\prime }=-\frac{1}{i
\rlap{\protect\rule[1.1ex]{.325em}{.1ex}}h%
\gamma ^{3/2}}[\hat{V},\hat{C}_{\square }^{eucl}(1)].  \label{18}
\end{equation}

Working out these expressions, the full constraint equation becomes 
\begin{equation}
(\Psi |(\hat{H}_{grav}^{(\mu _{0})}+16\pi G\hat{H}_{\phi })^{\dagger }=0;
\label{19}
\end{equation}
we shall use the expansion (7) for $(\Psi |$ and, with the help of equations
(15-18), rewrite (19) as a discrete evolution equation ( units c=1, 
\rlap{\protect\rule[1.1ex]{.325em}{.1ex}}h%
=1) (see [7], where such expression was for the first time derived): 
\[
\sqrt{\gamma }\frac{3}{l_{pl}^{2}\gamma ^{3/2}\mu _{0}^{3}}[(V_{\mu +5\mu
_{0}}-V_{\mu +3\mu _{0}})\psi _{\mu +4\mu _{0}}(\phi )-2(V_{\mu +\mu
_{0}}-V_{\mu -\mu _{0}})\psi _{\mu }(\phi )+ 
\]
\[
+(V_{\mu -3\mu _{0}}-V_{\mu -5\mu _{0}})\psi _{\mu -4\mu _{0}}(\phi )]- 
\]

\[
-2(1+\gamma ^{2})\frac{27}{2}\frac{1}{l_{pl}^{14}\gamma ^{9}\mu _{0}^{9}}%
[(V_{\mu +9\mu _{0}}-V_{\mu +7\mu _{0}})D_{\mu +8\mu _{0}}\psi _{\mu +8\mu
_{0}}(\phi )- 
\]
\[
-2(V_{\mu +\mu _{0}}-V_{\mu -\mu _{0}})D_{\mu }\psi _{\mu }(\phi )+(V_{\mu
-7\mu _{0}}-V_{\mu -9\mu _{0}})D_{\mu -8\mu _{0}}\psi _{\mu -8\mu _{0}}(\phi
)]= 
\]
\begin{equation}
=-16\pi G\hat{H}_{\phi }\psi _{\mu }(\phi ),  \label{20}
\end{equation}
where the coefficients $D_{\mu }$ are complicated functions of $V_{\mu }$
given in the appendix A. This equation constrains the functions $\psi _{\mu
}(\phi ),$ defined in equation (7), to ensure that the $(\Psi |$ belong to
the space of physical states. As mentioned in the introduction, the fact
that this equation is different from the corresponding equation in reference
[9], shows that the operations of reducing the phase space and quantizing do
not commute between them.

We used the ordering scheme adopted by Bojwald. Had we adopted a symmetrized
version, we could verify that $\psi _{\mu =0}$ would not decouple from the
system.

It is known that, in the case of constraint operators, we are only
interested in their kernel and the operators are not required to be
self-adjoint. This seems to be an advantage in quantum gravity, an open
algebra category of constrained system [13], as the self-adjoint requirement
may lead to quantum anomalies. On the other hand, dropping this requirement,
we are left with the following awkward situation. The matter hamiltonian is
required from quantum mechanics to be self-adjoint, and $\hat{H}_{\phi }$
indeed obeys this requirement; then, taking the adjoint of the evolution
equation, we recover the same $\hat{H}_{\phi }$, but not the same $\hat{H}%
_{grav}^{(\mu _{0})}$. We are left with two distinct choices, where we
should have one.

However, we are here interested in the large $\mu $ behaviour of the system
and it happens that both versions, symmetrized and non-symmetrized, lead to
exactly the same equations; we may study this behaviour without bothering
with this problem. To investigate the limit of large $\mu $, it is more
convenient to study the differential equation resulting from (20) in such a
limit.

Before proceeding, let us mention that, applying the methods developped in
[9], section 4.3, we again find the classical limit 
\begin{equation}
<H_{grav}^{(\mu _{0})}>=\frac{6}{l_{pl}^{2}\gamma ^{2}}c_{0}^{2}\sqrt{p}+...
\label{21}
\end{equation}
with $p\equiv \gamma l_{pl}^{2}n\mu _{0}/6$ , $n\gg 1$, where we added
contributions coming from the two terms in equation (20).

\bigskip

III. \underline{The large volume limit}.

It is important to keep present the fact that the large volume limit, large
values of $\mu $, it is not the same as the limit, usually taken, $\mu
_{0}\rightarrow 0.$ This can be seen, for instances, from inspection of
equation (42) in [9]. It is the behaviour in the large volume limit we are
interested in. Working out this limit in terms of $(\mu _{0}/\mu )$, all the
volume functions appearing on the l.h.s. of the evolution equation become
proportional to $\sqrt{\mu \pm n\mu _{0}}$ times powers of $\mu _{0}$. Then,
adding and subtracting, after multiplication by an appropriate constant, the
terms corresponding to the first bracket (in $\psi _{\mu \pm 4\mu _{0}}$ and 
$\psi _{\mu }$) of equation (20), we can rewrite the l.h.s. in a form
appropriate to this limit: 
\[
l.h.s.=\frac{1}{\gamma }-\frac{1+\gamma ^{2}}{\gamma ^{3}})\frac{9}{\mu
_{0}^{3}}[\frac{1}{\mu _{0}^{2}}(\sqrt{\mu +4\mu _{0}}\psi _{\mu +4\mu
_{0}}(\phi )-2\sqrt{\mu }\psi _{\mu }(\phi )+\sqrt{\mu -4\mu _{0}}\psi _{\mu
-4\mu _{0}}(\phi ))]- 
\]

\[
-\frac{9}{4}\frac{1+\gamma ^{2}}{\gamma ^{3}}\frac{1}{\mu _{0}}[\frac{1}{\mu
_{0}^{4}}(\sqrt{\mu +8\mu _{0}}\psi _{\mu +8\mu _{0}}(\phi )-4\sqrt{\mu
+4\mu _{0}}\psi _{\mu +4\mu _{0}}(\phi )+ 
\]

\begin{equation}
+6\sqrt{\mu }\psi _{\mu }(\phi )-4\sqrt{\mu -4\mu _{0}}\psi _{\mu -4\mu
_{0}}(\phi )+\sqrt{\mu -8\mu _{0}}\psi _{\mu -8\mu _{0}}(\phi ))].
\label{22}
\end{equation}

What we have, within the square brackets, are the algorithms for second and
fourth derivatives. The limit we are going to take is not $\mu
_{0}\rightarrow 0$, but the limit of large $\mu $; the replacement of the
discrete expressions by derivatives is only an approximation which becomes
better and better as $\mu $ gets larger and larger.

Converting the combinations of $\mu $, $\mu \pm 4\mu _{0}$, etc. into
momentum variables, through equations (4) and (5), we finally find the
equation replacing (20): 
\[
\frac{2}{3}l_{pl}^{4}[f(p)\sqrt{p}\psi (p,\phi )]^{\prime \prime }+\frac{2}{%
27}\gamma ^{2}(1+\gamma ^{2})\mu _{0}^{2}l_{pl}^{8}(\sqrt{p}\psi (p,\phi
))^{\prime \prime \prime \prime }\approx 
\]
\begin{equation}
\approx -16\pi G\hat{H}_{\phi }\psi (p,\phi ),  \label{23}
\end{equation}
where the derivatives are with respect to the momentum $p$. The term $f(p)$
is given by

\begin{equation}
f(p)=(1-\frac{(\mu _{0}\gamma l_{pl}^{2})^{2}}{288}\frac{1}{p^{2}}+\ldots )
\label{24}
\end{equation}
and is usually put equal to 1, neglecting the small corrections.

We see that, had we taken the limit $\mu _{0}\rightarrow 0$, we would have
recovered the standard form of the Wheeler-De Witt equation. This not being
the case, we find in addition an extra term involving a fourth derivative.
This term comes from the two-fold application, when deriving (20), of the $%
\hat{C}^{eucl}$ operator which appears through the definition of the
extrinsic curvature operator $\hat{K}^{\prime }$, in equation (18). Having
begun with an hamiltonian formulation of a classical theory involving second
derivatives, we end up with a quantum equation including a fourth derivative.

In the vacuum case, equation (23) has analytical solutions of the form found
by Bojowald in [7], with two of them obeying the pre-classicality condition
of mild variation. When the matter term is included, we have not found
solutions that could be put into an analytical form. We included the matter
term in the form of a scalar field, using the procedure outlined in section
II.a, equations (7 - 14). We solved equation (23) by numerical methods (see
figures 1-4, with $y=\sqrt{p}\psi $) and found that we continue to have
solutions, fig. 1 for instances, whose behaviour can be made as slow as we
wish, by a convenient choice of initial conditions. They thus seem to
fulfill the basic requirement for pre-classicality.

\bigskip

IV. \underline{The classical equations of motion}.

We would now like to ask which classical action, using the conventional
methods, gives us back the Wheeler-De Witt equation (23), without the term
in $(\sqrt{p}\psi )^{\prime \prime \prime \prime }$, and with $\hat{H}_{\phi
}$ defined by (14). (We follow J. B. Hartle's lectures, ref. [15]).

The restriction to a homogeneous and isotropic geometry gives the simplest
class of minisuperspace models. The line element is

\begin{equation}
ds^{2}=-N^{2}(t)dt^{2}+a^{2}(t)d\vec{x}^{2},  \label{25}
\end{equation}
where now $N(t)$ is the (arbitrary) lapse function and $a(t)$ the
dimensionless scale factor. We continue to take for the matter content a
massive scalar field and define our effective classical action by the
expression (representing, for the moment, the scalar field by $\varphi $)

\[
S(a,\varphi )=\frac{1}{2}\int dtN[\frac{1}{16\pi G}(-6\frac{1}{f(a)}\frac{1}{%
a}(\frac{aa^{\prime }}{N})^{2}+6ka)+\frac{1}{\overline{A}(a)}(\frac{\varphi
^{\prime }}{N})^{2}- 
\]

\begin{equation}
-\frac{1}{2}\overline{B}(a)m_{pl}^{2}m^{2}\varphi ^{2},  \label{26}
\end{equation}
where $m$ is in planck units, $(^{\prime })\equiv d/dt$ and the lagrangean
is given by $L=\delta S/\delta t$; from now on we put $k=0$, flat case. The
only non-conventional elements are the functions $\overline{A}(a)$ and $%
\overline{B}(a)$, to be defined later, and $f(a)$, given by (24), now
expressed as a function of the scale $a$. Introducing the momenta

\begin{equation}
\pi _{a}=\frac{\delta L}{\delta a^{\prime }}=-\frac{6}{16\pi G}\frac{%
aa^{\prime }}{Nf(a)}  \label{27}
\end{equation}

\begin{equation}
\pi _{\varphi }=\frac{\delta L}{\delta \varphi ^{\prime }}=\frac{1}{%
\overline{A}}\frac{\varphi ^{\prime }}{N},  \label{28}
\end{equation}
we find the constraint by varying $L$ with respect to $N$; writing the
resulting equation in terms of $\pi _{a}$ and $\pi _{\varphi }$ we have

\begin{equation}
\frac{16\pi G}{12}\frac{1}{a}\pi _{a}^{2}f(a)-\frac{1}{2}\overline{A}(a)\pi
_{\varphi }^{2}-\frac{1}{2}\overline{B}(a)m_{pl}^{2}\varphi ^{2}=0.
\label{29}
\end{equation}
(Note the positions of $\overline{A}(a)$ and $f(a)$ in (26) and (29)).

We shall now replace $\varphi $ by $\phi /\sqrt{6}$, followed by the usual
replacements of $\pi _{a}$ and $\pi _{\phi }$ by the operators

\begin{equation}
\pi _{a}\rightarrow -i(\frac{1}{16\pi G})^{3/2}\frac{\partial }{\partial a}
\label{30}
\end{equation}

\begin{equation}
\pi _{\phi }\rightarrow -i(\frac{1}{16\pi G})\frac{\partial }{\partial \phi }%
,  \label{31}
\end{equation}
where we took $\phi $ to be dimensionless. The factor ordering ambiguities,
affecting the gravitational operator, will be lifted by writing it in the
form suggested by lqc; then equation (29) becomes ($\frac{1}{2}%
m_{pl}^{2}\equiv \frac{1}{16\pi G}$)

\begin{equation}
\lbrack -\frac{1}{2}a(\frac{1}{a}\frac{\partial }{\partial a}\frac{1}{a}%
\frac{\partial }{\partial a}f(a))+\frac{1}{2}\overline{A}\frac{\partial ^{2}%
}{\partial \phi ^{2}}-\frac{1}{2}\overline{B}m^{2}\phi ^{2}]\Psi (a,\phi )=0,
\label{32}
\end{equation}
after dropping an overall constant factor $\frac{1}{6}(\frac{1}{16\pi G}%
)^{2} $.

To find the Wheeler-De Witt equation (23), all that remains is to express $a$
in terms of $p$ and introduce $A(p)$ and $B(p)$, the already known functions
given by equations (12) and (13), but now expressed directly in terms of $p$
($V=|p|^{3/2}$). Remembering that $a$ is dimensionless ($a^{2}$ is in fact
the $\mu $ of sections II and III), we introduce the relation

\begin{equation}
-\frac{1}{2}a(\frac{1}{a}\frac{\partial }{\partial a}\frac{1}{a}\frac{%
\partial }{\partial a})=-2(\frac{\gamma l_{pl}^{2}}{6})^{3/2}\sqrt{p}(\frac{%
\partial ^{2}}{\partial p^{2}})  \label{33}
\end{equation}
and the definitions

\begin{equation}
16\pi GA(p)\equiv \frac{1}{3}(\frac{6}{\gamma })^{3/2}l_{pl}\overline{A}(p)
\label{34}
\end{equation}

\begin{equation}
16\pi GB(p)\equiv \frac{1}{3}(\frac{6}{\gamma })^{3/2}l_{pl}\overline{B}(p)
\label{35}
\end{equation}
\begin{equation}
\Psi (p,\phi )\equiv \sqrt{p}\psi (p,\phi ).  \label{36}
\end{equation}
With the help of these expressions we find

\begin{equation}
\frac{2}{3}l_{pl}^{4}\frac{\partial ^{2}}{\partial p^{2}}(f(p)\sqrt{p}\psi
(p,\phi ))=16\pi G[\frac{1}{2}A(p)\frac{\partial ^{2}}{\partial \phi ^{2}}-%
\frac{1}{2}B(p)m^{2}\phi ^{2}]\psi (p,\phi ),  \label{37}
\end{equation}
which is equation (23) without the the term $(\sqrt{p}\psi )^{\prime \prime
\prime \prime }$ and with $\hat{H}_{\phi }$ given by (14). Thus, we may now
use the action $S(a,\varphi =\phi /\sqrt{6})$ to derive the effective
classical equations of motion (see Bojowald, ref. [16], for a different way
of arriving at these modified equations).

Again, varying $L=\delta S/\delta t$ with respect to $N$ we get the modified
Friedman's equation, after replacing $\overline{A}(a)$ and $\overline{B}(a)$
by $A(a)$ and $B(a)$:

\begin{equation}
\frac{1}{f(a)}(\frac{a^{\prime }}{a})^{2}=\frac{8\pi G}{3}\{\frac{1}{2}[%
\frac{1}{36(l_{pl}a)^{3}A(a)}(\frac{6}{\gamma })^{3/2}](\phi ^{\prime })^{2}+%
\frac{1}{2}[\frac{B(a)}{2(l_{pl}a)^{3}}(\frac{\gamma }{6})^{3/2}]m^{2}\phi
^{2}\}.  \label{38}
\end{equation}
We see that the form of the equation is correct. With $A(a)$ representing,
respectively, the inverse volume and the volume, apart from some constants
which can be absorbed into $\phi $ and $m$, we have that $a^{3}A(a)$ and $%
B(a)/a^{3}$become equal to1 in the large volume limit; we thus recover the
standard form of the equation, taking also into account that $f(a)\approx 1$.

Variation with respect to the scalar field gives

\begin{equation}
\phi ^{\prime \prime }-\frac{A(a)^{\prime }}{A(a)}\phi ^{\prime }+(18(\frac{%
\gamma }{6})^{3}A(a)B(a))m^{2}\phi =0.  \label{39}
\end{equation}
(See also Appendix B). Equations (38) and (39) are, with small differences, the equations normally
used in the classical limit, when we wish to include corrections derived
from lqc.

That is, from the same effective classical action we can derive both, the
modified Wheeler-De Witt equation (23) and the equations (38) and (39). 
\newline

\bigskip 

As for the term $\partial ^{4}/\partial p^{4}$, we have not
yet found a reasonable way of incorporating into the effective action a new
term corresponding to this fourth-order derivative. At first sight, we might
think that, at least in simple situations, such a term would give a
correction to the Friedamn's equation proportional to $(a^{\prime }/a)^{4}$,
but a lot of care is necessary. This would come out by adding a term
to $S$ of the form $ca(a^{\prime })^{4}/N^{3}$, $c$ proportional to the
constants multiplying $\partial ^{4}/\partial p^{4}$ in (23) (in planck units, $c$ is of the order of $\gamma^{2}\mu_{0}^{2}$). The new
momentum $\pi _{a}$ would become

\begin{equation}
\pi _{a}=(1-\frac{4}{3}cl_{pl}^{2}\frac{(a^{\prime })^{2}}{%
N^{2}})(-6\frac{aa^{\prime }}{16\pi GN})\simeq (-6\frac{aa^{\prime }}{16\pi
GN})  \label{40}
\end{equation}
in those, and only in those, situations where the second term in the first bracket can be
neglected, that is, when $(\gamma\mu_{0})^{2}l_{pl}^{2}(a^{\prime})^{2}\ll 1$. Then, varying $L$ with
respect to $N$, we find that (for simplicity, we put $f(a)=1$)
\begin{equation}
\frac{16\pi G}{12}\frac{1}{a}\pi _{a}^{2}\rightarrow \frac{16\pi G}{12}\frac{%
1}{a}\pi _{a}^{2}-\frac{3}{6^{4}}c(16\pi G)^{4}\frac{1}{a^{3}}\pi _{a}^{4}.
\label{41}
\end{equation}
After the replacement of $\pi _{a}$ defined in (30), we would get, as in (23), the extra fourth-order derivative term
$\sqrt{p} \partial ^{4}/\partial p^{4}$ (the factor $\sqrt{p}$ later cancelling out) and, in
the Friedman's equation, a new term proportional to $a^{2}(a^{\prime }/a)^{4}$. Corrections of
this form to the Friedman's equation do, sometimes, appear in the literature
of brane based cosmology ([19] and references therein).

\bigskip

V. \underline{Summary}.

In the present work, after deriving a modified form of the Wheeler-De Witt
equation, as the large volume limit of the discrete evolution equation of
loop quantum cosmology, we looked for an effective classical action from
which this equation could be obtained, using the conventional methods. Once
this action has been defined, we may also derive from it the associated
classical equations, the modified Friedman's equation and the scalar
equation of motion for the massive scalar field, which we took as our matter
content. These are the equations normally used in the applications of the
loop quantum cosmology to the study of inflation. Numerical simulations were
done with the modified Wheeler-De Witt equation, and we found that solutions
exist that comply with the requirements of pre-classicality. We also found
numerical solutions to the modified equation for the scalar field (Appendix B), having,
as expected, the appropriate inflationary behaviour.

Further applications will be left to a future work.

\bigskip

\bigskip \underline{Appendix A }

The functions $D_{\mu }$ appearing in equation (20) are given by the
following expressions:

\[
D_{\mu +8\mu _{0}}=D_{1}^{\prime }[(V_{\mu +6\mu _{0}}-V_{\mu +4\mu
_{0}})(V_{\mu +5\mu _{0}}-V_{\mu +\mu _{0}})-
\]
\begin{equation}
-(V_{\mu +4\mu _{0}}-V_{\mu +2\mu _{0}})(V_{\mu +3\mu _{0}}-V_{\mu -\mu
_{0}})]  \label{42}
\end{equation}

\[
D_{\mu }=\frac{1}{2}D_{1}[(V_{\mu -2\mu _{0}}-V_{\mu -4\mu _{0}})(V_{\mu
+\mu _{0}}-V_{\mu -3\mu _{0}})- 
\]

\[
+(V_{\mu -4\mu _{0}}-V_{\mu -6\mu _{0}})(V_{\mu -\mu _{0}}-V_{\mu -5\mu
_{0}})]+ 
\]

\[
+\frac{1}{2}D_{2}(V_{\mu +6\mu _{0}}-V_{\mu +4\mu _{0}})(V_{\mu +5\mu
_{0}}-V_{\mu +\mu _{0}})-
\]
\begin{equation}
-(V_{\mu +4\mu _{0}}-V_{\mu +2\mu _{0}})(V_{\mu +3\mu _{0}}-V_{\mu -\mu
_{0}})]  \label{43}
\end{equation}

\[
D_{\mu -8\mu _{0}}=D_{2}^{\prime }[(V_{\mu -2\mu _{0}}-V_{\mu -4\mu
_{0}})(V_{\mu +\mu _{0}}-V_{\mu -3\mu _{0}})-
\]
\begin{equation}
-(V_{\mu -4\mu _{0}}-V_{\mu -6\mu _{0}})(V_{\mu -\mu _{0}}-V_{\mu -5\mu
_{0}})],  \label{44}
\end{equation}
where $D_{1}$ and $D_{2}$ are

\[
D_{1}=[(V_{\mu +2\mu _{0}}-V_{\mu })(V_{\mu +\mu _{0}}-V_{\mu -3\mu _{0}})-
\]
\begin{equation}
-(V_{\mu }-V_{\mu -2\mu _{0}})(V_{\mu -\mu _{0}}-V_{\mu -5\mu _{0}})]
\label{45}
\end{equation}

\[
D_{2}=[(V_{\mu +2\mu _{0}}-V_{\mu })(V_{\mu +5\mu _{0}}-V_{\mu +\mu _{0}})-
\]
\begin{equation}
-(V_{\mu }-V_{\mu -2\mu _{0}})(V_{\mu +3\mu _{0}}-V_{\mu -\mu _{0}})]
\label{46}
\end{equation}
and 
\[
D_{1}^{\prime }=[(V_{\mu +10\mu _{0}}-V_{\mu +8\mu _{0}})(V_{\mu +9\mu
_{0}}-V_{\mu +5\mu _{0}})-\bigskip 
\]
\begin{equation}
-(V_{\mu +8\mu _{0}}-V_{\mu +6\mu _{0}})(V_{\mu +7\mu _{0}}-V_{\mu +3\mu
_{0}})]  \label{47}
\end{equation}

\[
D_{2}^{\prime }=[(V_{\mu -6\mu _{0}}-V_{\mu -8\mu _{0}})(V_{\mu -3\mu
_{0}}-V_{\mu -7\mu _{0}})-
\]
\begin{equation}
-(V_{\mu -8\mu _{0}}-V_{\mu -10\mu _{0}})(V_{\mu -5\mu _{0}}-V_{\mu -9\mu
_{0}})]\bigskip .  \label{48}
\end{equation}

\bigskip

\underline{Appendix B}

Let us apply a simple formal procedure, to obtain equation (39). Assume a
regime where it is valid to use the commutators

\[
\lbrack \hat{\phi},\hat{p}_{\phi }]=i
\rlap{\protect\rule[1.1ex]{.325em}{.1ex}}h%
, 
\]

\begin{equation}
d\hat{p}_{\phi }/dt=\frac{1}{i
\rlap{\protect\rule[1.1ex]{.325em}{.1ex}}h%
}[\hat{p}_{\phi },\hat{H}_{\phi }]  \label{49}
\end{equation}
and 
\begin{equation}
d\hat{\phi}/dt=\frac{1}{i
\rlap{\protect\rule[1.1ex]{.325em}{.1ex}}h%
}[\hat{\phi},\hat{H}_{\phi }],  \label{50}
\end{equation}
followed by a discretization, through the replacement of $dt$ by a discrete
step parameter $h$. We get the equations ($\hat{H}_{\phi }$ given by (14)) 
\begin{equation}
\phi (\mu +h)=\phi (\mu )+hA_{\mu }p_{\phi }(\mu )  \label{51}
\end{equation}

\begin{equation}
p_{\phi }(\mu +h)=p_{\phi }(\mu )+hB_{\mu }m^{2}\phi (\mu ).  \label{52}
\end{equation}
A little algebra shows that we can derive the following equation for $\phi $:

\[
\frac{1}{h^{2}}(\phi (\mu +2h)-2\phi (\mu +h)+\phi (\mu ))=(\frac{1}{A_{\mu }%
}\frac{A_{\mu +h}-A_{\mu }}{h})\frac{1}{h}(\phi (\mu +h)-\phi (\mu ))- 
\]

\begin{equation}
-A_{\mu +h}B_{\mu }m^{2}\phi (\mu )  \label{53}
\end{equation}
which, again, suggests that, in this limit, we should replace the $3(\dot{a}%
/a)$ term by the expression within the first bracket on the r.h.s. Numerical
integration of this equation, for a large variety of initial conditions,
with $h\sim 4\mu _{0}$, shows that, after a period of slow increase, $\phi $
goes through a period of very fast increase until it settles down into an
almost constant value, a behaviour that is well known and is the starting
point of quantum geometry inflation [17, 18]. A simple numerical example can
be seen in figure 5.

\bigskip

\underline{Acknowledgements} - I would like to thank Jos\'{e} Velhinho for
his many suggestions and help in calculations.

 This work was supported by FCT, Funda\c{c}\~{a}o para a Ci\^{e}ncia e a Tecnologia, Portugal, under the project POCTI/FIS/57547/2004.

\bigskip

\pagebreak

REFERENCES

[1] M. Bojowald, Class. Quant. Grav. \textbf{17} (2000), 1489; gr-qc/9910103.

[2] M. Bojowald, Class. Quant. Grav. \textbf{18} (2001), 1071; gr-qc/0008053.

[3] M. Bojowald, Phys. Rev. Lett. \textbf{86} (2001), 5227; gr-qc/0102069.

[4] M. Bojowald, Phys. Rev. Lett.$\Bbb{\ }$\textbf{87} (2001), 121301;
gr-qc/0104072.

[5] M. Bojowald, Phys. Rev. \textbf{D64} (2001), 084018; gr-qc/0105067.

[6] M. Bojowald, Class. Quant. Grav. \textbf{18} (2001), L109; gr-qc/0105113.

[7] M. Bojowald, Class. Quant. Grav. \textbf{19} (2002), 2717; gr-qc/0202077.

[8] M. Bojowald, Phys. Rev. Lett. \textbf{89} (2002), 261301; gr-qc/0206054.

[9] A. Ashtekar, M. Bojowald and J. Lewandowski, Adv. Theor. Math. Phys. 
\textbf{7 }(2003), 233; gr-qc/0304074.

[10] J. Velhinho, Class. Quant. Grav. \textbf{21} (2004), L109;
gr-qc/0406008.

[11] H. Nicolai, K. Peeters and M. Zamaklar, Class. Quant. Grav. \textbf{22}
(2005), R193; hep-th/0501114 v1.

[12] T. Thiemann, Class. Quant. Grav. \textbf{15} (1998), 1281;
gr-qc/9705019.

[13] T. Thiemann, ``Introduction to Modern Canonical Quantum General
Relativity'', pre-print gr-qc/0110034.

[14] A. Ashtekar, J. Lewandowski, Class. Quant. Grav. \textbf{21} (2004)
R53; gr-qc/0404018.

[15] James B. Hartle, ''Quantum Cosmology'', Lectures delivered at the
Theoretical Advanced Study Institute in Elementary Particle Physics, Yale
University, June 10-14, 1985; preprint 1985.

[16] M. Bojowald, Pramana \textbf{63} (2004), 765; gr-qc/0402053.

[17] M. Bojwald, H. A. Morales-T\'{e}cotl, Lect. Notes Phys. \textbf{646}
(2004) 421; gr-qc/0306008.

[18] M. Bojowald, J.E. Lidsey, D. J. Mulryne, P. Singh and R. Tavakol, Phys.
Rev. \textbf{D70} (2004), 043530; gr-qc/0403106.

[19] G. Dvali and M. S. Turner, ``Dark Energy as a Modification of the
Friedman Equation'', astro-ph/0301510.

\pagebreak

\bigskip

\bigskip

\includegraphics{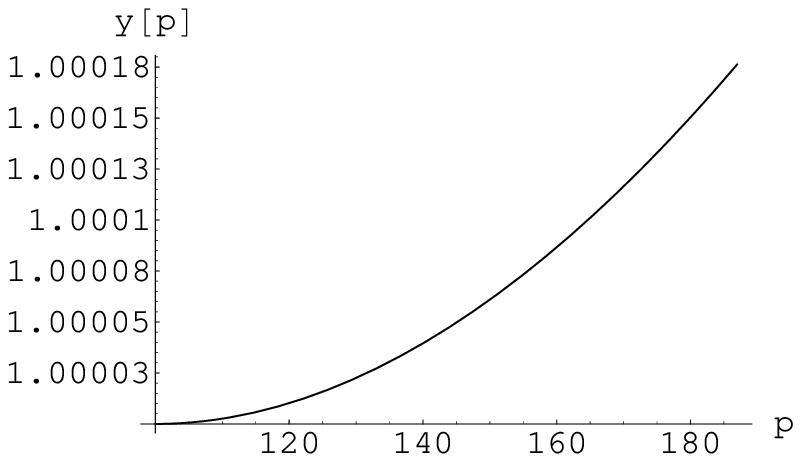}

Fig. 1 - The graph shows the solution of equation (23), the initial
conditions are $y^{\prime \prime \prime }=y^{\prime \prime }=y^{\prime }=0$, 
$y=1$ at $p_{0}=100$. We assumed a scalar field mass equal to 0.1 planck
units.

\bigskip

\bigskip

\bigskip

\bigskip

\includegraphics{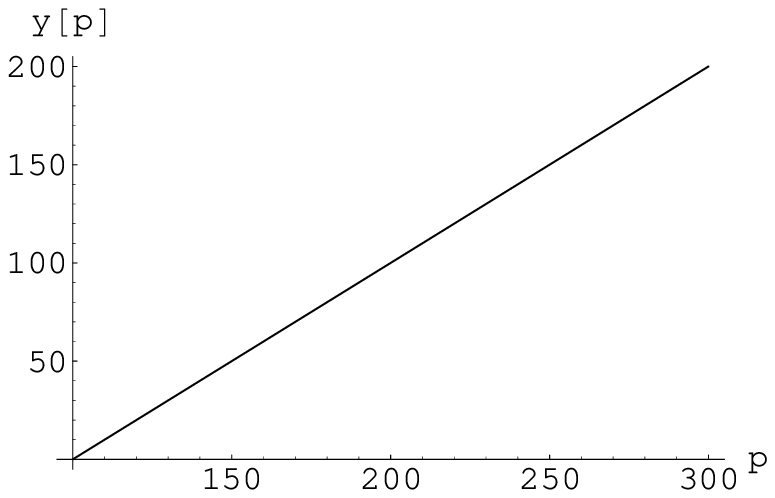}

Fig. 2 - The same as figure 1, except that now $y^{\prime }=1$ and $%
y^{\prime \prime \prime }=y^{\prime \prime }=y=0$, at $p_{0}=100$.

\pagebreak

\bigskip

\bigskip

\includegraphics{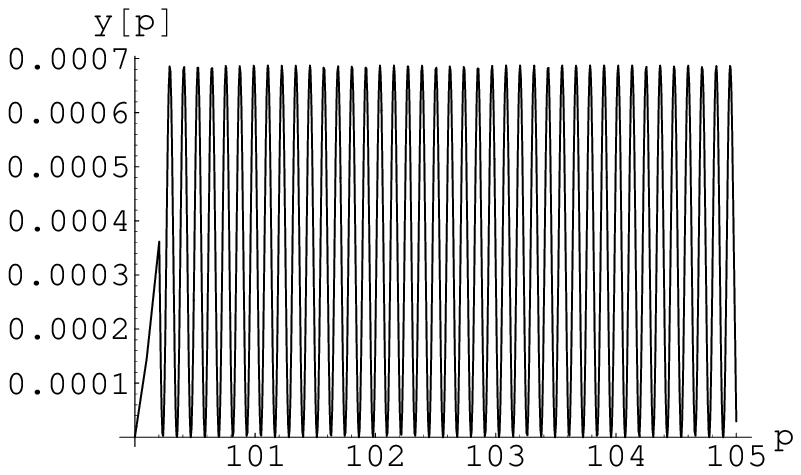}

Fig. 3 - Solution of equation (23); initial conditions $y^{\prime \prime }=1$%
, $y^{\prime \prime \prime }=y^{\prime }=y=0$, at $p_{0}=100$.

\bigskip

\bigskip

\bigskip

\bigskip

\includegraphics{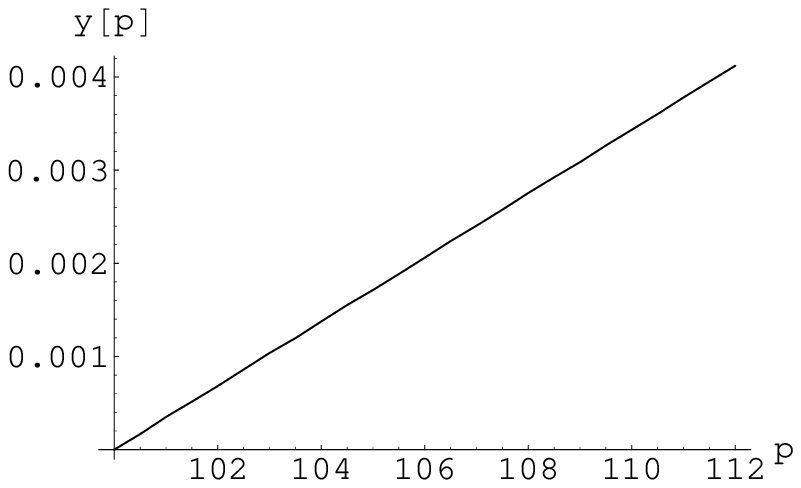}

Fig. 4 - The same as previous figures; initial conditions $y^{\prime \prime
\prime }=1$, $y^{\prime \prime }=y^{\prime }=y=0$, at $p_{0}=100$.

\pagebreak

\bigskip

\bigskip

\includegraphics{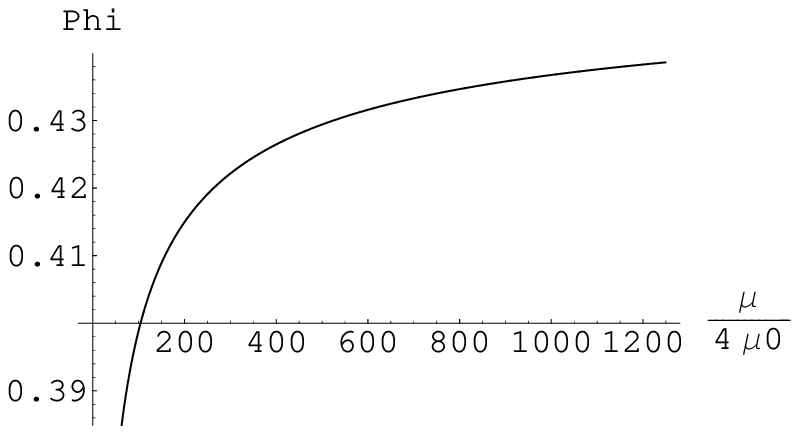}

Fig. 5 - Numerical integration of the discrete equation (53) for the scalar
field. Initial conditions were $\phi (0)=0$, $\phi ^{\prime }(0)=0.1$,
scalar field mass=0.1 planck units.

\end{document}